\documentclass[11pt]{article}
\usepackage{jheppub}
\usepackage{tikz}
\usetikzlibrary{
    calc,
    arrows.meta, 
    patterns,      
    decorations.pathreplacing,  
    shapes.misc 
}
\usepackage{ulem}
\usepackage[T1]{fontenc}
\usepackage{amssymb}
\usepackage{mathtools}
\usepackage{chngcntr}
\usepackage{stackengine}
\usepackage{scalerel}

\usepackage{hyperref}
\usepackage{epsfig}
\usepackage{amsmath,latexsym,amssymb}
\usepackage{graphicx}
\usepackage{braket}
\usepackage{pdfsync}

\usepackage{framed}
\usepackage{mathtools}

\usepackage{jheppub}
\usepackage[T1]{fontenc}
\usepackage{amssymb}
\usepackage{mathtools}
\usepackage{amsmath}
\usepackage{bm}
\usepackage{stackengine}

\usepackage{scalerel}
\usepackage{eufrak}
\usepackage{framed}

\usepackage{mathtools}

\usepackage{pstricks}

\usepackage{amsmath}
\usepackage{bbold}
\usepackage{bm}
\usepackage{latexsym}
\usepackage{braket}
\usepackage{slashed}
\usepackage{graphicx,booktabs,multirow}
\usepackage{eufrak}

\usepackage{color}
\usepackage{pstricks}
\usepackage{amssymb}

\makeatletter
\newcommand{\doublewidetilde}[1]{{%
  \mathpalette\double@widetilde{#1}%
}}
\newcommand{\double@widetilde}[2]{%
  \sbox\z@{$\m@th#1\widetilde{#2}$}%
  \ht\z@=.9\ht\z@
  \widetilde{\box\z@}%
}
\makeatother

\usepackage{hyperref}
\usepackage{slashed}
\usepackage{epsfig}
\usepackage{amsmath,latexsym,amssymb}
\usepackage{graphicx}
\usepackage[latin1]{inputenc}
\usepackage{braket}
\usepackage{pdfsync}

\usepackage{framed}

\usepackage{mathtools}

\def\be{\begin{equation}}
\def\ee{\end{equation}}
\def\ba{\begin{eqnarray}}
\def\ea{\end{eqnarray}}

\renewcommand{\j}{\mathsf{3j}}

\def\k{\kappa}

\def\L{\Lambda}

\newcommand{\comment}[1]{}

\newcommand{\eea}{\end{eqnarray}}

\def\Tr{{\rm Tr}}

\counterwithout{equation}{section}

\renewcommand{\thesection}{\arabic{section}}
\author{
Tomasz R.\ Taylor${}^{1,2}$,\, Bin Zhu${}^{3}$\\[0.5cm]
 $^1${\it Department of Physics,
  Northeastern University, Boston, MA 02115, USA}\\
  $^2${\it Faculty of Physics, University of Warsaw, ul. Pasteura 5, 02-093 Warsaw, Poland}\\
$^3${\it School of Physics, Nankai University, Weijin Road 94, Tianjin 300071, P.R. China}\\[0.2cm]
}

\emailAdd{taylor@neu.edu}
\emailAdd{bzhu@nankai.edu.cn}

\title{ \hskip 2 cm Three-Gluon Scattering  Amplitude\\ \hskip 4cm  in de Sitter Spacetime}

\abstract{\hfill\\ We study three-gluon scattering amplitudes in global de Sitter spacetime, in the angular momentum basis of $SO(1,4)$ symmetry representations. At the tree level, they are determined by the intertwiner integrals of harmonic one-forms on the three-sphere. We derive a general formula valid for all helicity configurations of incoming and outgoing gluons and express the amplitudes in terms of Wigner $\j$ symbols.

\vskip 3 cm
\centerline{\it Dedicated to the memory of Jerzy Lewandowski}}
\makeatletter
\gdef\@fpheader{}
\makeatother

\begin{document}
\maketitle
\section{Introduction}
Three-gluon  scattering amplitudes are the fundamental constructs and  basic building blocks of perturbative Yang-Mills theory \cite{Taylor:2017sph}. In this work, we study three-gluon amplitudes in curved spacetime.
In the physical universe, the curvature varies from place to place, depending on the matter and radiation densities. Here, we focus on de Sitter spacetime because constant curvature allows a tractable case study. Assuming maximal symmetry, the S-matrix formalism can be  generalized  to {\it global\/} de Sitter spacetime \cite{Taylor:2024vdc,Taylor:2025deepIR}.\footnote{For earlier work, see Refs.\ \cite{Marolf:2012kh,Melville:2024ove}.}

In flat spacetime, free particle states form unitary representations of the Poincar\'e group. In a similar way, de Sitter particles form representations of the $SO(1,4)$ de Sitter isometry group \cite{Dixmier:1961fbm,rep4,Lindsay:2025chz}. A remarkable feature of the representations associated to the particles of the standard model and gravitons is the absence of the ``zero energy modes'' which are responsible for the  infrared divergences of scattering amplitudes in flat spacetime \cite{Lindsay:2025chz}. The curvature ``regulates'' such divergences. For that reason, global de Sitter spacetime may serve not only as a model of our physical universe, but also as a framework for addressing the notorious problem of infrared divergences.

\section{Quantization}
De Sitter spacetime has the topology of $\mathbb{R}\times S^3$. We are considering Yang-Mills theory in the gravitational background of de Sitter metrics \cite{rep21}. At the perturbative level, Yang-Mills theory is a theory of interacting gluons --  quantum excitations  of a  Lie algebra valued one-form gauge field $A$. We further restrict to the semi-classical approximation, in which Yang-Mills theory is conformally invariant and the background metric is equivalent to the metric on a strip of Einstein cylinder:
\begin{equation}
ds^2=-dt^2+d\Omega^2 ,
\end{equation}
where the time coordinate $t\in (-\pi/2,\pi/2)$ and $ d\Omega^2$ is the metric on $S^3$. The Yang-Mills action  integral over de Sitter volume is:
\be
{\cal S}=\frac{1}{2g^2}\int \Tr (F\wedge\!*F)\ ,\qquad F=dA+A\wedge A\ ,\label{saction}
\ee
where $F$ is the Yang-Mills field strength two-form, $*$ denotes the Hodge dual and  $g$ is the coupling constant.

To quantize the gauge field, we solve the wave equation and identify the positive and negative frequency modes. In the Coulomb gauge, the solutions  factorize as \be A=e^{\mp i \omega t} E\ ,\ee where $\omega\ge 0$ and $E$ are one-forms in the bundle cotangent to $S^3$. The one-forms associated to massless gauge bosons are eigenforms of the Hodge-Laplace operator on $S^3\,$:
\be (\delta d+d\delta)E=k^2E\, ,~~\makebox{with}~ k=\omega\, \label{hl}
\ee
and are coexact (transverse), {\it i.e}.\ \be\delta E=0\  ,\ee where $\delta$ is the codifferential. 
Hereafter, these wavefunctions  are called transverse covector harmonics.

The wavefunctions describe one-gluon states of the Hilbert space. De Sitter Hilbert space can be decomposed as the sum $\oplus_{(j, j')\in \Gamma}{\cal H}_{j,j'}$ of  $SU(2)\times SU(2)'$ representations ${\cal H}_{j,j'}$,  $(j,j')\in \Gamma$, with the set $\Gamma$ depending on the  SO(1,4) representation content \cite{Dixmier:1961fbm}. The  $SU(2)\times SU(2)'$ ``isospin'' group emerges from the isometry of de Sitter's  $S^3$ spatial sections.
The basis of ${\cal H}_{j,j'}$ consists of 
\be  
\big|j\,\nu\rangle|j'\nu'\big\rangle
\ ,\quad\nu=-j,-j+1,\dots,j ,~~\nu'=-j',-j'+1,\dots,j'\ ,\label{dst}\ee
with  the  $SU(2)\times SU(2)'$ isospins  $j\ge 0$ and $j'\ge0$ taking half-integer values.
Massless gauge bosons belong to the  unitary irreducible representations in the discrete series, $\Pi^+_1$ and $\Pi^-_1$, characterized by the Casimir operator values corresponding to the positive and negative helicities  $\pm
1$ (right- and left-handed polarizations), respectively \cite{rep21,Lindsay:2025chz}.\footnote{When $SO(1,4)$ is considered as the conformal symmetry group, these representations are also characterized by the conformal dimension $\Delta=2$.}  On the $(j,j')$ plane shown in Figure 1, these representations populate straight lines with
\be j'=j+1~~\makebox{for}~ \Pi^+_1\ ,~~~~~~~~~~j'=j-1~~\makebox{for}~ \Pi^-_1\ .\ee
\begin{figure}[t]
\centering
\begin{tikzpicture}[scale=0.9]
\draw[step=0.5, thin, color=gray!20] (0,0) grid (5,5);
\draw[->] (0,0) -- (5,0) node[right] {$j$};
\draw[->] (0,0) -- (0,5) node[above] {$j'$};
\draw[dash pattern=on 3pt off 1pt, line width=0.4pt, red] (0,1) -- (4,5);
\draw[dash pattern =on 3pt off 1pt, line width=0.4pt, blue] (1,0) -- (5, 4);
\node[left] at (0,1) {$1$};
\node[below] at (1,0) {$1$};
\foreach \k in {0,...,8}{\fill[red] (\k*0.5, \k*0.5 + 1) circle (2pt);}
\foreach \k in {0,...,8}{\fill[blue] (\k*0.5+1, \k*0.5) circle (2pt);}
\node[red] at (2,3.7,0) {${\Pi^+_1}$}; 
\node[blue] at (3.9,2) {${\Pi^-_1}$};
\node[left] at (0,0.5) {$\frac{1}{2}\,$};
\node[below] at (0.5,0) {$\frac{1}{2}$};
\node[left] at (0,1) {$1$};
\node[below] at (1,0) {$1$};
\end{tikzpicture}
\caption{Isospin content of  $\Pi^+_1$ and $\Pi^-_1$ representing polarized gluons in de Sitter spacetime \cite{Lindsay:2025chz}.}
\end{figure}
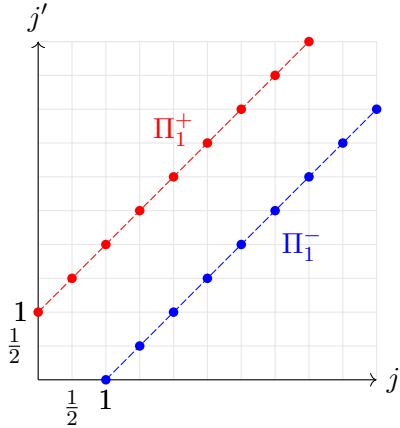
It is convenient to use one integer, $k\ge 2$, to label  the $SU(2)\times SU(2)'$ multiplets,
\be k=2j'=2j+2~~ \makebox{for}~ \Pi^+_1\ , ~~~~~~~   k=2j=2j'+2~~ \makebox{for}~ \Pi^-_1\ .\label{kdef}\ee

Transverse covector harmonics have integer eigenvalues $k^2$ of the Hodge-Laplace operator (\ref{hl}), with $k\ge2$.  They fall into two classes, right- and left-handed, with the helicity determined by the sign on the r.h.s.\ of
\be *dE^\pm=\pm kE^\pm\ .\label{heli}\ee
{}For each $k$,  they  transform linearly under the $SU(2)\times SU(2)'$ isometry. They match all  irreducible representations of $SU(2)\times SU(2)'$ contained in $ \Pi^+_1$ and $ \Pi^-_1$, in one-to-one correspondence with the gluon states. These wavefunctions intertwine the cotangent bundle over  $S^3$  with de Sitter Hilbert space. Note that the frequencies $\omega=k\ge 2$ therefore, as mentioned before, the zero frequency modes are absent.

\section{Scattering amplitudes}
According to the formalism developed in  Refs.\cite{Taylor:2024vdc,Taylor:2025deepIR}, three-gluon amplitudes are 
determined at the tree level by the overlap of  the wavefunctions in the cubic part of the action integral  (\ref{saction}):
\be \langle out|in\rangle_{3}={ig}\sum\int \Tr(A_{a_1}\wedge A_{a_2}\wedge *dA_{a_3})\ ,\ee
where the sum is over the permutations  of $a_1,a_2,a_3$ labelling three sets of gluon quantum numbers. 
{}For incoming particles, we insert the wave functions; for outgoing particles, we use the corresponding complex conjugates.

\subsection{All-plus amplitudes}
First, we consider  ``all plus'' amplitude describing  three incoming right-handed gluons disappearing into the vacuum. After using the helicity constraints (\ref{heli}) and taking the trace, we obtain
\begin{align}{\cal M}(1^+_{in}2^+_{in}3^+_{in})&=-igf_{a_1a_2a_3} (k_1+k_2+k_3)\int  e^{-it(\omega_1+\omega_2+\omega_3)}dt\wedge E^+_1\wedge E^+_2\wedge E^+_3\nonumber\\[1mm] & \label{m123}
= -2igf_{a_1a_2a_3} {\sin[(\omega_1+\omega_2+\omega_3)\frac{\pi}{2}]\over \omega_1+\omega_2+\omega_3}(k_1+k_2+k_3)\, I_3\ ,\end{align}
where $f_{a_1a_2a_3}$ are the Lie algebra structure constants. The  intertwiner integral is
\be I_3=\int_{S^3}E^+_{(j_1,\nu_1)(j'_1,\nu'_1)}\wedge E^+_{(j_2,\nu_2)(j'_2,\nu'_2)}\wedge E^+_{(j_3,\nu_3)(j'_3,\nu'_3)}\ .\label{intt}\ee

\subsection{Intertwiner}
In order to compute the intertwiner (\ref{intt}), we use the Hopf angles as (toroidal) coordinates  on $S^3$: 
 $\chi \in [0, \pi/2]$, $\theta\in [0, 2\pi) $,  $ \varphi\in [0, 2\pi)$ \cite{Lehoucq:2002E,Lindsay:2025chz}. Then the metric reads
\begin{equation}
d\Omega^2 = d\chi^2 + \cos^2{\chi} d\theta^2 + \sin^2{\chi}d\varphi^2.\label{s3}
\ee
Explicit expressions for the covector harmonics \cite{BenAchour:2015aah} can be obtained from the solutions of the Laplace equation
\begin{equation}
\Delta Y_{klm}(\Omega)  = -k(k+2)Y_{klm}(\Omega) \, .
\end{equation}
The scalar harmonics $Y_{klm}$ are labelled by integer $k\ge 0$, with the integers $l$ and $m$
constrained by $|m|+|l|\le k$ and $|m|+|l|= k$ mod 2. They are given by \cite{Lehoucq:2002E,Lindsay:2025chz}:
\begin{equation}
Y_{klm}(\Omega)=\frac{i^{|m|}}{i^m} \frac{1}{2\pi}\sqrt{\frac{2(k+1) d!(|l|+|m|+d)!}{(|l|+d)!(|m|+d)!}} \, \cos^{|l|}\!\chi \sin^{|m|}\!\chi P_{d}^{|m|,|l|}(\cos 2\chi) e^{il\theta} e^{im\varphi} \, , \label{eq:Yklm}
\end{equation}
where  $P_{d}^{|m|,|l|}$ are the Jacobi polynomials of degree $d=\frac{1}{2}(k-|m|-|l|) $. 
The  $SU(2)\times SU(2)'$ quantum numbers are identified as
\be
j=\frac{k}{2}-1\ , ~~j'=\frac{k}{2}\ , ~~\nu=\frac{l+m}{2}\ , ~~\nu'=\frac{l-m}{2}\ ,\ee
with $k\ge 2$ and $\nu,\nu'$ constrained by (\ref{dst}). Then the right-handed harmonics
\be E^+_{(j,\nu)(j',\nu')}={\cal N}_{klm}^+\,(E^+_\chi d\chi+E^+_\theta d\theta +E^+_\varphi d\varphi)\ , \ee
where  the components are given by
\begin{align}
E_\chi^+ &= i(l+m)\partial_\chi Y_{klm}+ik(l\tan\chi-m\cot\chi)Y_{klm} \, ,\\[1mm]
E_\theta^+ &=-k\sin\chi\cos\chi \partial_\chi Y_{klm} +(k^2\cos^2\chi -l(l+m))Y_{klm} \, , \\[1mm]
E_\varphi^+ &=k\sin\chi \cos\chi \partial_\chi Y_{klm} +(k^2\sin^2\chi-m(l+m))Y_{klm} \, .
\end{align}
Imposing the normalization condition
\be \int_{S^3}E^*\wedge *E=k\ ,\ee
where $E^*$ is the complex conjugate of $E$, yields the corresponding normalization factor
\be
\mathcal{N}^+_{klm}= \frac{1}{\sqrt{2(k+1)(k^2-(l+m)^2)}}\ .
\ee

The intertwiner integral (\ref{intt}) is invariant under   $SU(2)\times SU(2)'$ $S^3$ isometry. For $SU(2)$, there is a unique invariant tensor connecting three representations. It is the Wigner 
$\mathsf{3j}$ symbol \cite{varsh}, which is related to Clebsch-Gordan coefficients in the following way:
\be \left( {j_1\atop n_1}  {j_2\atop n_2}  {j_3\atop n_3}\right)=\frac{(-1)^{j_1-j_2- n_3}}{\sqrt{2j_3+1}}
\langle j_1 n_1,j_2 n_2|j_3\,{-} n_3\rangle\ .\ee
According to the rules of quantum angular momentum addition, it vanishes unless the triangle conditions and the magnetic selection rule are fulfilled:
\be |j_1-j_2|\le j_3\le j_1+j_2~~~\makebox{and}~~~n_1+n_2+n_3=0\ .\ee
Hence the $SU(2)\times SU(2)'$ symmetry fixes the integral (\ref{intt}), up to an $SU(2)\times SU(2)'$ invariant overall coefficient (depending on $j$ and $j'$ only), to be proportional to a product of two Wigner $\j$ symbols. Evaluating this coefficient, we obtain
\be
I_3=\frac{i\mathsf{S}}{\pi}\sqrt{s^2-1}
\left( {j_1\atop\nu_1}\,  {j_2\atop\nu_2} \, {j_3\atop\nu_3}\right)
\left( {j'_1\atop\nu'_1}\,  {j'_2\atop\nu'_2} \, {j'_3\atop\nu'_3}\right)\ ,\label{ires}\ee
where 
\be  \mathsf{S}=\sqrt{s(s-k_1)(s-k_2)(s-k_3)}\ ,\qquad s=\frac{k_1+k_2+k_3}{2}\ .\ee
In Appendix~\ref{app:intertwiner-derivation}, we derive the intertwiner (\ref{ires}) explicitly for the nontrivial family of modes satisfying $l+m=0$ and verify it independently for general admissible mode configurations.
Note that $\mathsf{S}$ is the area of a triangle with the side lengths $(k_1,k_2,k_3)$. Recall that in the right-handed case, $k=2j'$.\footnote{From the mathematical perspective, there is nothing more natural than to integrate the exterior product of three harmonic one-forms over $S^3$, therefore we expect that Eq.(\ref{ires}) can be rederived by using abstract representation theory.} 

In general, incoming and outgoing gluons can be right- or left-handed. The intertwiner integrals involve not only right- but also left-handed covector harmonics. Their components are given by
\begin{align}
E^-_\chi &= i(l-m)\partial_\chi Y_{klm}+ik(l\tan\chi+m\cot\chi)Y_{klm} \, ,\\
E^-_\theta &=-k\sin\chi\cos\chi \partial_\chi Y_{klm} +(k^2\cos^2\chi -l(l-m))Y_{klm} \, , \\
E^-_\varphi &=-k\sin\chi \cos\chi \partial_\chi Y_{klm} -(k^2\sin^2\chi-m(m-l))Y_{klm} \, .
\end{align}
and the normalization factor
\be
\mathcal{N}^-_{klm}= \frac{1}{\sqrt{2(k+1)(k^2-(l-m)^2)}}\ .
\ee
Note that complex conjugation does not change the helicity, but only flips the angular momentum components $l$ and $m$ to $-l$ and $-m$. Recall that in the left-handed case, $k=2j$. After performing integrations, one finds that replacing one or more $E^+$ by 
$E^-$ in the intertwiner integral $I_3$ (\ref{intt}) amounts to the replacement of the respective $k$ by $-k$ in Eq.(\ref{ires}).

\subsection{All three-gluon amplitudes}
At this point, we can write down a compact formula for all three-gluon amplitudes. We assign the numbers
\be
{\lambda=+1 ~~ \makebox{helicity}\, +\atop \lambda=-1  ~~\makebox{helicity}\, -}\qquad\qquad
{\epsilon=+1 ~~ \makebox{incoming} \atop \epsilon=-1  ~~\makebox{outgoing} }\ee
and define
\be \lambda_k=\frac{\lambda_1k_1+\lambda_2k_2+\lambda_3k_3}{2}\ ,\qquad \epsilon_k=\frac{\epsilon_1k_1+\epsilon_2k_2+\epsilon_3k_3}{2}\ .\ee
After combining all intermediate expressions and taking into account the dispersion relation $\omega=k$, we obtain
\be
{\cal M}(1^{\lambda_1}_{\epsilon_1}\, 2^{\lambda_2}_{\epsilon_2}\, 3^{\lambda_3}_{\epsilon_3})= 2gf_{a_1a_2a_3}\,\lambda_k\sqrt{\lambda_k^2-1}\,\mathsf{S}\left( {j_1\atop\epsilon_1\nu_1}\,  {j_2\atop\epsilon_2\nu_2} \, {j_3\atop\epsilon_3\nu_3}\right)
\left( {j'_1\atop\epsilon_1\nu'_1}\,  {j'_2\atop\epsilon_2\nu'_2} \, {j'_3\atop\epsilon_3\nu'_3}\right) \frac{\sin\epsilon_k\pi }{\epsilon_k\pi}\  .\label{result}
\ee
It follows from the relation of wavenumbers $k\ge 2$  to the angular momentum numbers, see Eq.(\ref{kdef}), and from the triangle condition, that $\epsilon_k$  and $\lambda_k$ are integer numbers different from zero, therefore all three-gluon amplitudes vanish due to the presence of  the ``energy-conserving'' sine factor.
This is not unexpected. In flat spacetime, the energy-momentum conservation law forces the four-momenta of three massless  particles  to a  collinear configuration, in which all Lorentz-invariant kinematic variables become zero. Hence the kinematics are overconstrained and all three-gluon amplitudes are zero \cite{Taylor:2017sph}. 
Once the particle momenta are extended from real to complex values, however, they exhibit beautiful holomorphic and antiholomorphic structures. In de Sitter background, the structures that emerge from intertwiners are equally interesting. They are related to representation theory and will  reappear in higher-point amplitudes. For that reason, we
discuss below the properties of amplitudes that do not depend on the kinematic constraints. They are:\\
$\quad${\it ~1}. Bose symmetry: the amplitudes remain unchanged when two incoming or outgoing gluons with identical helicities are interchanged. Since $j'=j\pm 1$, the product of $\j$ symbols acquires a minus sign when two gluons are interchanged, which is compensated by the antisymmetry of structure constants.\\
$\quad${\it ~2}. Time reversal $(\epsilon_1,\epsilon_2,\epsilon_3)\to (-\epsilon_1,-\epsilon_2,-\epsilon_3)$:  this symmetry, which holds up to the sign of amplitude, moves all particles from initial to final state  and vice versa.\\
$\quad${\it ~3}. Parity $(\lambda_1,\lambda_2,\lambda_3)\to(-\lambda_1,-\lambda_2,-\lambda_3)$: this symmetry, which holds up to the sign of amplitude, flips signs of all gluon helicities.

De Sitter scattering amplitudes match their flat counterparts in the limit of large angular momentum
\cite{Taylor:2024vdc,Taylor:2025deepIR,Lindsay:2025chz}. For three gluons, both amplitudes vanish due to the kinematic constraints. We expect that for more gluons, the large angular momentum limit will allow studying the limit of small curvature and a comparison with the flat limit.
\section{Amplitudes vs cosmological correlators}
In inflationary cosmology, the Poincar\'e patch of de Sitter manifold, {\it a.k.a}.\ flat slicing of de Sitter hyperboloid, provides a mathematical model of  rapid expansion of early universe. In this framework, the spectrum (and eventual non-Gaussianities) of  primordial density fluctuations can be related to the ``cosmological correlators'' of quantum fields in de Sitter background
\cite{Maldacena:2002vr,Maldacena:2011nz,Arkani-Hamed:2015bza,Chen:2017ryl,MdAbhishek:2025dhx,Arundine:2026fbr}.
The cosmological correlators are evaluated at a fixed cosmological time, while the scattering amplitudes  describe quantum fields evolving from far past to the future.
They are related,  however, as elucidated in Ref.\cite{Chen:2017ryl}. It is known that the cosmological correlators do not vanish in the case of three gluons \cite{Maldacena:2011nz,MdAbhishek:2025dhx,Arundine:2026fbr}, which appears to contradict the vanishing of three-gluon amplitudes discovered  in the present work.

At the technical level, the main difference between the calculations of amplitudes and cosmological correlators is the time integration. In the cosmological case, the integral runs over the Schwinger-Keldysh contour and requires a $\pm i\epsilon$ prescription. In the present case of three-gluon amplitude, the time integration is trivial, {\it c.f.}\ Eq.(\ref{m123}):
\be \int_{-\pi/2}^{\pi/2}  e^{-it(\omega_1+\omega_2+\omega_3)}dt=2 {\sin[(\omega_1+\omega_2+\omega_3)\frac{\pi}{2}]\over \omega_1+\omega_2+\omega_3}.\ee
The time interval is the length of the Poincar\'e cylinder, and the integration is convergent without any prescription. This ``energy-conserving'' sine factor is responsible for the vanishing of the amplitudes. Is there a symmetry reason for this result? In four dimensions, Yang-Mills theory is conformally invariant at the tree level -- which justifies replacing de Sitter manifold by the Poincar\'e cylinder.  In principle, without conformal invariance, time integration could yield a different result and give rise to nonvanishing three-gluon amplitudes. To clarify this point,  in Appendix B we compute the amplitudes in the presence of  $F^3$ interactions that violate conformal invariance. We show that the global de Sitter time integral is finite and the resulting three-gluon amplitudes are zero again. Most likely, the vanishing of all  three-gluon amplitudes follows from the full $SO(1,4)$ symmetry, in a way similar to the vanishing of ``all out'' scalar amplitudes explained in Ref.\cite{Lindsay:2025chz} by using group-theoretical arguments.\footnote{The reason is that the product of three principal series does not contain a trivial representation. Group-theoretical aspects of gluon scattering will be discussed in a future work.} On the other hand, the cosmological correlators of gauge fields are evaluated on the Poincar\'e patch, with the gluon states labeled by continuous three-momentum. There is no direct connection between such states and the global de Sitter $SO(1,4)$ discrete series. Although both cosmological correlators and amplitudes are constrained by $SO(1,4)$ Ward identities, the amplitudes are subject to additional constraints due to the properties of discrete series representations \cite{Lindsay:2025chz}.

\section{Summary and outlook}
In this work, we expressed three-gluon amplitudes in the angular momentum basis, in terms of Wigner $\j$ symbols. These amplitudes will serve as building blocks for perturbative Yang-Mills theory in de Sitter spacetime. It is clear that the Feynman rules will be similar to the graphical theory of angular momentum \cite{varsh,graph}. We expect more $SU(2)\times SU(2)'$ invariant structures to appear in multi-gluon amplitudes, in particular the cel{\nolinebreak}ebrated  $6\mathsf{j}$ symbols \cite{six}.\footnote{Similar structures appear in  the transition amplitudes of spin foam theory of loop quantum gravity which, affording bold speculation,  may signal spacetime ``emergent'' from the scattering amplitudes.}  
It would be also interesting to extend the result (\ref{result}) to complex angular momentum, to obtain expressions that could be compared with the (nonvanishing) complexified three-gluon amplitudes in flat spacetime.

In global de Sitter spacetime, massless gauge bosons (and gravitons) have no zero energy/frequency modes. The gluons with lowest angular momentum $(k=2)$ have the wavefunctions determined by the six Killing vectors of $S^3$. Three-gluon vertices involving such ``soft'' gluons have a particularly simple form, therefore it should not be difficult to investigate how the so-called soft theorems generalize to curved spacetime. In the framework of celestial holography,  soft theorems lead to a rich structure encapsulated in the infinite-dimensional $\mathcal{L}w_{1+\infty}$ symmetry algebra \cite{Strominger:2021mtt}. One important question that could be answered after generalizing soft theorems to de Sitter spacetime, is whether they are related to the recently discovered ``cosmologically-deformed'' $\mathcal{L}_{\Lambda}w_{1+\infty}$ algebra \cite{Taylor:2023ajd,Bittleston:2024rqe}.

 Another interesting aspect of de Sitter Yang-Mills theory is the clear separation of positive and negative helicity states into two distinct representations of the $SO(1,4)$ isometry group. Hence the ``on-shell helicity formalism'' is already in place.\footnote{See also Refs.\cite{Maldacena:2011nz,Basile:2024ydc}.} In the flat limit, when the spacetime is extended to complex and/or split signature spaces, single helicity sectors originate from self-dual or anti-self dual sectors of gauge theory.  We hope that de Sitter  becomes a fruitful arena for studying dynamics of self-dual gauge theories.

\section*{Acknowledgements}
\addcontentsline{toc}{section}{Acknowledgements}
We would like to thank Nima Arkani-Hamed, Laurent Freidel and Guilherme Pimentel for enlightening conversations during the 2026 annual meeting of the Simons Collaboration on Celestial Holography in New York. We are grateful to Akshay Venkatesh and Griffin Wang for sharing with us their insights into the mathematics of intertwiners.
This work was supported in part by NSF PHY-2209903, the Simons Collaboration on Ce{\nolinebreak}lestial Holography, and by the MAESTRO grant no.\ 2024/54/A/ST2/00009
funded by the National Science Centre, Poland. It was also supported by
Polish National Agency for Academic Exchange under the NAWA Chair programme.
Any opinions, findings, and conclusions or
recommendations expressed in this material are those of the authors and do not necessarily
reflect the views of the National Science Foundation. Bin Zhu is supported by the Fundamental Research Funds for the Central Universities (010-63263123).

\appendix
\addtocontents{toc}{\protect\setcounter{tocdepth}{1}}
\numberwithin{equation}{section}

\section{Analytic derivation of the three-form intertwiner}
\label{app:intertwiner-derivation}

The $SU(2)\times SU(2)'$ covariance of the spatial integral fixes the general
intertwiner in Eq.~(\ref{ires}) up to a coefficient.  We determine this
coefficient by a direct calculation for positive-helicity modes with $l+m=0$.
We then generalize the result to arbitrary helicities and incoming or outgoing
configurations, as shown in Eq.~(\ref{result}).  We also perform
checks for general mode configurations.

\subsection{Conventions and normalization}

We use the toroidal coordinates $x^i=(\chi,\theta,\varphi)$ and the unit
three-sphere metric given in Eq.~(\ref{s3}), with
$0\leq\chi\leq\pi/2$ and $0\leq\theta,\varphi<2\pi$.
The orientation and volume form are
\begin{equation}
 d\chi\wedge d\theta\wedge d\varphi>0,
 \qquad
 \mathrm{vol}_{S^3}=\sin\chi\cos\chi\,
 d\chi\wedge d\theta\wedge d\varphi .
 \label{eq:appA-orientation}
\end{equation}
The scalar harmonics are unit normalized, whereas the one-form harmonics used
in the main text obey
\begin{equation}
 \int_{S^3}Y_{klm}^{*}Y_{k'l'm'}\,\mathrm{vol}_{S^3}
 =\delta_{kk'}\delta_{ll'}\delta_{mm'},
 \qquad
 \int_{S^3}(E^\lambda_{klm})^*\wedge *E^{\lambda'}_{k'l'm'}
 =k\,\delta_{\lambda\lambda'}\delta_{kk'}
       \delta_{ll'}\delta_{mm'} .
 \label{eq:appA-paper-normalization}
\end{equation}
All one-form harmonics and intertwiner integrals below use this main-text
normalization.  

\subsection{The \texorpdfstring{$l+m=0$}{l+m=0} modes}

We begin with the scalar harmonic in Eq.~(\ref{eq:Yklm}), including its phase
$i^{|m|}/i^m$, and set $z=\cos(2\chi)$.  When $l+m=0$, the two indices of the
Jacobi polynomial coincide.  The identity
\begin{equation}
 P_n^M(z)=(-1)^M\frac{(n+M)!}{2^M n!}
 (1-z^2)^{M/2}P_{n-M}^{(M,M)}(z).
 \label{eq:appA-jacobi-legendre}
\end{equation}
shows that it can be replaced by an associated Legendre function.  For
$M=|l|=|m|$ and $n=k/2$, the complete phase multiplying that function is
therefore
\begin{equation}
 \frac{i^{|m|}}{i^m}(-1)^M.
 \label{eq:appA-combined-phase}
\end{equation}
We consider the following three positive-helicity modes:
\begin{align}
 (k_1,l_1,m_1)&=(2L_1,L_1,-L_1),
 & (k_2,l_2,m_2)&=(2L_2,-L_1,L_1),
 & (k_3,l_3,m_3)&=(2L_3,0,0),
 \label{eq:appA-labels}
\end{align}
where $L_1,L_2,L_3$ are positive integers and $L_2\geq L_1$.  In the phase
convention of the main text, the corresponding scalar harmonics are
\begin{align}
 Y_1={}&\frac{1}{2\pi}
 \sqrt{\frac{2(2L_1+1)}{(2L_1)!}}\,
 P_{L_1}^{L_1}(z)e^{iL_1(\theta-\varphi)},
 \label{eq:appA-Y1}\\
 Y_2={}&\frac{(-1)^{L_1}}{2\pi}
 \sqrt{\frac{2(2L_2+1)(L_2-L_1)!}{(L_2+L_1)!}}\,
 P_{L_2}^{L_1}(z)e^{-iL_1(\theta-\varphi)},
 \label{eq:appA-Y2}\\
 Y_3={}&\frac{1}{2\pi}\sqrt{2(2L_3+1)}\,
 P_{L_3}(z).
 \label{eq:appA-Y3}
\end{align}
The relative sign $(-1)^{L_1}$ in $Y_2$ is fixed by the phase in
Eq.~(\ref{eq:Yklm}) together with the factor $(-1)^M$ in
Eq.~\eqref{eq:appA-jacobi-legendre}; the corresponding factors in $Y_1$ and
$Y_3$ are unity.  Thus no additional phase convention is introduced in this
specialization.  For any $l+m=0$ mode, the positive-helicity component formulas
in the main-text intertwiner subsection reduce to the coordinate-basis
one-form
\begin{equation}
 E^+_{klm}=\frac{1}{k\sqrt{2(k+1)}}
 \begin{pmatrix}
 ik\bigl(l\tan\chi-m\cot\chi\bigr)Y_{klm}\\[2pt]
 -k\sin\chi\cos\chi\,\partial_\chi Y_{klm}
       +k^2\cos^2\!\chi\,Y_{klm}\\[2pt]
 k\sin\chi\cos\chi\,\partial_\chi Y_{klm}
       +k^2\sin^2\!\chi\,Y_{klm}
 \end{pmatrix}_{(d\chi,d\theta,d\varphi)} .
 \label{eq:appA-Eplus}
\end{equation}
In particular, $m=-l$ turns the first entry into
$ilY_{klm}/[\sqrt{2(k+1)}\sin\chi\cos\chi]$.  Direct substitution verifies
$\delta E^+_{klm}=0$, where $\delta=-*d*$.  Together with Eqs.~(\ref{hl}) and
(\ref{heli}), this confirms that Eq.~\eqref{eq:appA-Eplus} is the transverse
positive-helicity mode used in the main text.  Its prefactor also gives the
one-form normalization in Eq.~\eqref{eq:appA-paper-normalization}.

\subsection{Direct evaluation of the three-form}

Writing $E_a=E_{a\chi}d\chi+E_{a\theta}d\theta+E_{a\varphi}d\varphi$, we first
separate the angular and radial integrations.  The coefficient of the oriented
coordinate volume form is the determinant
\begin{equation}
 E_1\wedge E_2\wedge E_3
 =R_{L_1L_2L_3}(\chi)\,
 d\chi\wedge d\theta\wedge d\varphi,
 \qquad
 R_{L_1L_2L_3}(\chi)=
 \det\!\begin{pmatrix}
 E_{1\chi}&E_{1\theta}&E_{1\varphi}\\
 E_{2\chi}&E_{2\theta}&E_{2\varphi}\\
 E_{3\chi}&E_{3\theta}&E_{3\varphi}
 \end{pmatrix}.
 \label{eq:appA-wedge-determinant}
\end{equation}
The angular phases in Eqs.~\eqref{eq:appA-Y1}--\eqref{eq:appA-Y3} cancel, as
required by the two magnetic selection rules.  The remaining determinant
depends only on $\chi$.  It can be reduced with
\begin{align}
 (L_3+1)P_{L_3+1}(z)
 &=(2L_3+1)zP_{L_3}(z)-L_3P_{L_3-1}(z),
 \label{eq:appA-legendre-rec1}\\
 (L_2-L_1+1)P_{L_2+1}^{L_1}(z)
 &=(2L_2+1)zP_{L_2}^{L_1}(z)
   -(L_2+L_1)P_{L_2-1}^{L_1}(z),
 \label{eq:appA-legendre-rec2}\\
 P_L^L(z)&=(-1)^L(2L-1)!!(1-z^2)^{L/2}.
 \label{eq:appA-edge-legendre}
\end{align}
The terms proportional to $zP_{L_3}(z)P_{L_2}^{L_1}(z)$ cancel between the
three rows of the determinant.  This cancellation removes the derivatives and
leaves the radial kernel
\begin{align}
 R_{L_1L_2L_3}(\chi)={}&
 \frac{iL_1L_3(L_1+L_2)}{2\pi^3}
 \sqrt{\frac{(L_2-L_1)!}{(2L_1)!(L_1+L_2)!}}
 (2L_1-1)!!\,\sin^{L_1-1}(2\chi)
 \nonumber\\
 &\times\Bigl[
 P_{L_3}(z)P_{L_2-1}^{L_1}(z)
 -P_{L_3-1}(z)P_{L_2}^{L_1}(z)
 \Bigr].
 \label{eq:appA-general-radial}
\end{align}
Equation~\eqref{eq:appA-general-radial} is valid for the full family in
Eq.~\eqref{eq:appA-labels}.  We now specialize to $L_2=L_1\equiv L$.  In this
subfamily the first two scalar harmonics saturate
$|l|+|m|=k$ and form an extremal pair; this is the sense in which we refer to
them as edge modes.  Since $P_{L-1}^{L}=0$, the radial kernel reduces to
\begin{equation}
 R_{L,L_3}(\chi)=
 \frac{i(-1)^{L+1}L^2L_3}{4^L\pi^3}
 \binom{2L}{L}\sin^{2L-1}(2\chi)
 P_{L_3-1}\bigl(\cos2\chi\bigr).
 \label{eq:appA-edge-radial}
\end{equation}
Consequently,
\begin{equation}
 I_3^{(L,L,L_3)}=(2\pi)^2
 \int_0^{\pi/2}R_{L,L_3}(\chi)\,d\chi .
 \label{eq:appA-edge-integral}
\end{equation}
The angular integrations give the factor $(2\pi)^2$.  With
$y=\cos2\chi$, the remaining integral follows from
\begin{equation}
 \int_{-1}^{1}(1-y^2)^{\mu-1}P_\nu(y)\,dy
 =\frac{\pi\Gamma(\mu)^2}
 {\Gamma(\mu+\nu/2+1/2)\Gamma(\mu-\nu/2)
  \Gamma(\nu/2+1)\Gamma(1/2-\nu/2)}.
 \label{eq:appA-legendre-integral}
\end{equation}
Taking $\mu=L$ and $\nu=L_3-1$ gives
\begin{equation}
 I_3^{(L,L,L_3)}=
 \frac{i(-1)^L4^{1-L}(2L)!}
 {\Gamma\!\left(L+\frac{1-L_3}{2}\right)
  \Gamma\!\left(L+\frac{L_3}{2}\right)
  \Gamma\!\left(\frac{L_3+1}{2}\right)
  \Gamma\!\left(-\frac{L_3}{2}\right)}.
 \label{eq:appA-gamma-result}
\end{equation}
The factor $(-1)^L$ in Eq.~\eqref{eq:appA-gamma-result} follows from the phase
$i^{|m|}/i^m$ in Eq.~(\ref{eq:Yklm}) and the relative sign in
Eq.~\eqref{eq:appA-Y2}.  It is essential for agreement with the two-$3j$
expression in Eq.~(\ref{ires}); for example, $L=L_3=1$ gives
$I_3=2i/\pi$ in both calculations.

The denominator of Eq.~\eqref{eq:appA-gamma-result} also makes the selection
rules manifest:
\begin{equation}
 L_3\ \hbox{odd},
 \qquad 1\leq L_3\leq 2L-1.
 \label{eq:appA-selection}
\end{equation}
For even $L_3$, the Legendre polynomial in
Eq.~\eqref{eq:appA-edge-radial} is odd under
$\chi\mapsto\pi/2-\chi$, so the integral vanishes.  Values
$L_3>2L-1$ are excluded by the triangle rule.  Thus the coordinate integral
reproduces both the value and the selection rules expected from the
representation-theoretic result.

\subsection{Comparison with the two-\texorpdfstring{$3j$}{3j} formula}

For the extremal subfamily, the $SU(2)\times SU(2)'$ labels of the three
positive-helicity modes are
\begin{align}
 (j_1,j'_1;\nu_1,\nu'_1)&=(L-1,L;0,L),
 \nonumber\\
 (j_2,j'_2;\nu_2,\nu'_2)&=(L-1,L;0,-L),
 \nonumber\\
 (j_3,j'_3;\nu_3,\nu'_3)&=(L_3-1,L_3;0,0).
 \label{eq:appA-rep-labels}
\end{align}
The general formula in Eq.~(\ref{ires}) therefore predicts
\begin{align}
 I_3^{(3j)}={}&\frac{iL_3}{\pi}
 \sqrt{(2L-L_3)(2L+L_3)
       \bigl[(2L+L_3)^2-1\bigr]}
 \nonumber\\
 &\times
 \begin{pmatrix}L-1&L-1&L_3-1\\0&0&0\end{pmatrix}
 \begin{pmatrix}L&L&L_3\\L&-L&0\end{pmatrix}.
 \label{eq:appA-two3j}
\end{align}
The extremal $3j$ symbol is
\begin{equation}
 \begin{pmatrix}L&L&J\\L&-L&0\end{pmatrix}
 =\frac{(2L)!}{\sqrt{(2L-J)!(2L+J+1)!}}.
 \label{eq:appA-extremal3j}
\end{equation}
For the other symbol, set $a=L-1$, $c=L_3-1$, and $g=a+c/2$.  The selection
rule in Eq.~\eqref{eq:appA-selection} implies that $c$ is even and
$c\leq2a$, in which case
\begin{equation}
 \begin{pmatrix}a&a&c\\0&0&0\end{pmatrix}
 =(-1)^g\frac{g!}{(c/2)!^2(a-c/2)!}
 \sqrt{\frac{(c!)^2(2a-c)!}{(2a+c+1)!}}.
 \label{eq:appA-zero3j}
\end{equation}
Substitution of Eqs.~\eqref{eq:appA-extremal3j} and
\eqref{eq:appA-zero3j} into Eq.~\eqref{eq:appA-two3j}, followed by the gamma
duplication and reflection identities, reproduces
Eq.~\eqref{eq:appA-gamma-result}, including its phase.  The coordinate-space
calculation therefore proves Eq.~(\ref{ires}) analytically for every allowed
$L$ and $L_3$ in the extremal subfamily, not only for isolated examples.

To check the formula beyond the Legendre-polynomial simplification, we also
evaluated 50 nonvanishing all-positive-helicity intertwiners constructed from
the generic Jacobi modes in Eq.~(\ref{eq:Yklm}).  None of these configurations
has $l+m=0$ for all three states.  The sample covers the wavenumber triples
$(3,3,2)$, $(3,3,4)$, $(4,4,2)$, and $(3,4,5)$.  Two representative results are
\begin{align}
 I_3\bigl[(3,-2,1),(3,-1,0),(4,3,-1)\bigr]
 &=\frac{4i}{\pi},
 \nonumber\\
 I_3\bigl[(3,-2,1),(4,-2,0),(5,4,-1)\bigr]
 &=\frac{3i}{\pi}\sqrt{\frac52}.
 \label{eq:appA-generic-checks}
\end{align}
and agree with Eq.~(\ref{ires}).  All 50 comparisons are exact, including the
phase and normalization.  They complement the analytic extremal-family proof
by probing the general Jacobi-polynomial sector.

The extension to arbitrary helicities and incoming or outgoing states requires
two further conventions.  First, Eq.~(\ref{eq:Yklm}) and the corresponding
one-form components imply
\begin{equation}
 \bigl(E^\lambda_{klm}\bigr)^*
 =(-1)^m E^\lambda_{k,-l,-m}.
 \label{eq:appA-conjugation-phase}
\end{equation}
We use the rephased dual basis
\begin{equation}
 E^\lambda_{\mathrm{out}}(k,l,m)
 \equiv(-1)^m\bigl(E^\lambda_{klm}\bigr)^*
 =E^\lambda_{k,-l,-m}.
 \label{eq:appA-outgoing-dual}
\end{equation}
for an outgoing state.  If literal complex-conjugate wavefunctions are used
instead, the reduced intertwiner acquires the overall phase
\begin{equation}
 \eta_{\mathrm{out}}
 =(-1)^{\sum_{r:\,\epsilon_r=-1}m_r}.
 \label{eq:appA-outgoing-total-phase}
\end{equation}
Second, the replacement $k_r\mapsto\lambda_r k_r$ in the reduced intertwiner
fixes a signed branch of the square root.  The explicit one-forms give
\begin{equation}
 \left[\sqrt{\lambda_k^2-1}\right]_{\{\lambda_r\}}
 =\eta_\lambda\left|\sqrt{\lambda_k^2-1}\right|,
 \qquad
 \eta_\lambda=(-1)^{N_-(N_--1)/2},
 \qquad
 N_-=\#\{r:\lambda_r=-1\}.
 \label{eq:appA-helicity-branch}
\end{equation}
The square root in Eq.~(\ref{result}) is understood in the signed sense
defined above.  The absolute value denotes its nonnegative magnitude, while
$\eta_\lambda$ supplies the helicity-dependent sign.  Although this sign is invisible
after the final sine factor sets the on-shell amplitude to zero, it is fixed
unambiguously by the nonzero spatial intertwiner.

We tested these conventions before imposing the final kinematic zero.  For
each configuration, the spatial integral was compared with
\begin{align}
 \int_{S^3}E^{\lambda_1}_{1,\mathrm{ext}}\wedge
 E^{\lambda_2}_{2,\mathrm{ext}}\wedge
 E^{\lambda_3}_{3,\mathrm{ext}}
 ={}&\eta_{\mathrm{out}}\eta_\lambda
 \frac{i\mathsf S}{\pi}\sqrt{\lambda_k^2-1}
 \nonumber\\
 &\times
 \begin{pmatrix}j_1&j_2&j_3\\
 \epsilon_1\nu_1&\epsilon_2\nu_2&\epsilon_3\nu_3\end{pmatrix}
 \begin{pmatrix}j'_1&j'_2&j'_3\\
 \epsilon_1\nu'_1&\epsilon_2\nu'_2&\epsilon_3\nu'_3\end{pmatrix},
 \label{eq:appA-general-spatial-check}
\end{align}
where $E_{r,\mathrm{ext}}=E_r$ for an incoming state and
$E_{r,\mathrm{ext}}=E_r^*$ for an outgoing state.  On the other hand, the time
integral is given by
\begin{equation}
 \int_{-\pi/2}^{\pi/2}
 e^{-iKt}\,dt
 =\pi\frac{\sin(\epsilon_k\pi)}{\epsilon_k\pi},
 \qquad
 K=\sum_r\epsilon_r k_r=2\epsilon_k.
 \label{eq:appA-time-check}
\end{equation}
The two factors were then assembled into the general amplitude in
Eq.~(\ref{result}), by
using the dual basis
\eqref{eq:appA-outgoing-dual} and the branch
\eqref{eq:appA-helicity-branch}.  The 150 exact configurations cover all eight
helicity patterns and all eight incoming/outgoing patterns.  Every selected
spatial coefficient is nonzero, and
Eq.~\eqref{eq:appA-conjugation-phase} was also verified for the 49 distinct
outgoing modes in the sample.  All spatial, time-integral, and assembled
amplitude comparisons agree exactly.

\section{Three-gluon amplitudes from the \texorpdfstring{$F^3$}{F3} interaction}
\label{app:f3-interaction}

We now examine whether the vanishing of the on-shell three-gluon amplitudes
persists after adding the parity-even dimension-six interaction
\begin{equation}
 S_{F^3}=\frac{\kappa}{\Lambda^2}
 \int d^4x\sqrt{-g}\,f^{abc}
 F^a_{\mu}{}^{\nu}F^b_{\nu}{}^{\rho}F^c_{\rho}{}^{\mu}.
 \label{eq:appB-action}
\end{equation}
Only the linearized field strength from each external leg contributes to the
cubic vertex.  We compute three-point amplitudes obtained by inserting the three
external field strengths in the order displayed in
Eq.~\eqref{eq:appB-action}. We show that the corresponding three-point amplitudes also vanish.


\subsection{Weyl factor and external modes}

We use the conformal global coordinates introduced in section 2.  We denote the physical de Sitter metric by $g_{\mu\nu}$ and
the Einstein-cylinder metric by $\widehat g_{\mu\nu}$.  Setting the de Sitter
radius to one, $\ell=1$, the two metrics are related by
$g_{\mu\nu}=\cos^{-2}t\,\widehat g_{\mu\nu}$.
The field strength with both indices lowered is independent of the Weyl
factor.  Each of the three raised indices in the contraction contributes a
factor $\cos^2t$, whereas the measure contributes $\cos^{-4}t$.  Hence
\begin{align}
 \sqrt{-g}\,
 F_\mu{}^\nu F_\nu{}^\rho F_\rho{}^\mu
 & =\cos^2t\sqrt{-\widehat g}\,
 \widehat F_\mu{}^\nu\widehat F_\nu{}^\rho
 \widehat F_\rho{}^\mu .
 \label{eq:appB-Weyl-factor}
\end{align}
The residual factor $\cos^2t$ distinguishes the $F^3$ interaction from
the conformally invariant Yang--Mills kinetic term in four dimensions and is
responsible for the shifted time channels derived below.

We also use the temporal-Coulomb gauge and the mode factorization of that
section.  For each leg, $\epsilon_r=+1$ denotes an incoming state and
$\epsilon_r=-1$ an outgoing state, while $E_{r,\mathrm{ext}}=E_r$ or $E_r^*$,
respectively, as in Appendix A.  Thus
$A^{(r)}=e^{-i\epsilon_r k_r t}E_{r,\mathrm{ext}}$.  Complex conjugation does
not change the helicity sign $\lambda_r$.

\subsection{Reduction of the Lorentz contraction}

Let $(E_{r,\mathrm{ext}})_{\hat a}$ denote components in an oriented
orthonormal frame on $S^3$, with $\epsilon_{\hat1\hat2\hat3}=+1$.  Here $K$
is defined in Eq.~\eqref{eq:appA-time-check}.  Using the transverse condition
$\delta E_r=0$ and
the helicity equation (\ref{heli}), a direct contraction of the three
linearized field strengths gives
\begin{align}
 \operatorname{tr}_{\mathrm L}\!\left(F^{(1)}F^{(2)}F^{(3)}\right)
 ={}&e^{-iKt}k_1k_2k_3\,
 \mathcal C(\epsilon_r,\lambda_r)\,
 \epsilon^{\hat a\hat b\hat c}
 (E_{1,\mathrm{ext}})_{\hat a}(E_{2,\mathrm{ext}})_{\hat b}
 (E_{3,\mathrm{ext}})_{\hat c},
 \label{eq:appB-local-contraction}\\
 \mathcal C(\epsilon_r,\lambda_r)={}&
 \lambda_1\lambda_2\lambda_3
 \left(1+\sigma_1\sigma_2+\sigma_1\sigma_3
              +\sigma_2\sigma_3\right),
 \qquad
 \sigma_r=\epsilon_r\lambda_r.
 \label{eq:appB-C-factor}
\end{align}
Here $\operatorname{tr}_{\mathrm L}$ is the trace over Lorentz
indices.\footnote{It is distinct from the color trace $\Tr$ used in the
Yang--Mills action in the main text.}
The orthonormal determinant in Eq.~\eqref{eq:appB-local-contraction} is the
coordinate coefficient of the triple wedge product divided by
$\sqrt\gamma=\sin\chi\cos\chi$:
\begin{equation}
 \epsilon^{\hat a\hat b\hat c}
 (E_{1,\mathrm{ext}})_{\hat a}(E_{2,\mathrm{ext}})_{\hat b}
 (E_{3,\mathrm{ext}})_{\hat c}
 =\frac{1}{\sqrt\gamma}
 \left(E_{1,\mathrm{ext}}\wedge E_{2,\mathrm{ext}}\wedge
 E_{3,\mathrm{ext}}\right)_{\chi\theta\varphi}.
 \label{eq:appB-frame-wedge}
\end{equation}
Thus, after integration over $S^3$, the Lorentz contraction contains precisely
the intertwiner $I_3$ defined in Eq.~(\ref{intt}); it need not be evaluated
again.  All dependence on incoming or outgoing assignment and helicity is
contained in $\mathcal C$.

Since every $\sigma_r$ equals $\pm1$, the bracket in
Eq.~\eqref{eq:appB-C-factor} is four when the three signs agree and zero
otherwise.  Hence
\begin{equation}
 \mathcal C(\epsilon_r,\lambda_r)=
 \begin{cases}
 4\lambda_1\lambda_2\lambda_3,
 &\sigma_1=\sigma_2=\sigma_3,\\[2pt]
 0,&\text{otherwise}.
 \end{cases}
 \label{eq:appB-chirality-rule}
\end{equation}
The variables $\sigma_r=\epsilon_r\lambda_r$ label the chiralities of the
linearized field strengths.  Mixed chiralities therefore vanish locally,
before either the time or the spatial integral is performed.

\subsection{Finite global-time integral}

The Weyl factor in Eq.~\eqref{eq:appB-Weyl-factor}, together with the explicit
time dependence in Eq.~\eqref{eq:appB-local-contraction}, produces
\begin{equation}
	 {\cal I}_K=\int_{-\pi/2}^{\pi/2}
 \cos^2t\,e^{-iKt}\,dt.
 \label{eq:appB-time-definition}
\end{equation}
Expanding $\cos^2t=(2+e^{2it}+e^{-2it})/4$ and integrating the three
exponentials gives
\begin{equation}
 {\cal I}_K=
 \begin{cases}
 \pi/2,&K=0,\\
 \pi/4,&K=\pm2,\\[2pt]
 \dfrac{4\sin(\pi K/2)}{K(4-K^2)},&K\neq0,\pm2.
 \end{cases}
 \label{eq:appB-time-result}
\end{equation}
The two $3j$ symbols in the spatial intertwiner require
$k_1+k_2+k_3$ to be even.  Since $\epsilon_r=\pm1$, one has
$K\equiv k_1+k_2+k_3\pmod 2$, and the allowed values of $K$ are therefore
even.  Equation~\eqref{eq:appB-time-result} can then be nonzero only in the
three channels
\begin{equation}
 K=0,\pm2.
 \label{eq:appB-time-channels}
\end{equation}

\subsection{Integrated kernel and vanishing theorem}

Combining the local contraction, the Weyl factor, and both integrations gives
the ordered color-dressed kernel
\begin{equation}
 \mathcal K_{F^3}(1,2,3)=
 \frac{\kappa}{\Lambda^2}
 f^{a_1a_2a_3}k_1k_2k_3\,
 \mathcal C(\epsilon_r,\lambda_r)\,
 {\cal I}_K\,I_3,
 \label{eq:appB-full-kernel}
\end{equation}
where the intertwiner is the same object as in Appendix A, now evaluated on
the external one-forms:
\begin{equation}
 I_3=
 \int_{S^3}E_{1,\mathrm{ext}}\wedge E_{2,\mathrm{ext}}
 \wedge E_{3,\mathrm{ext}}.
 \label{eq:appB-signed-I3}
\end{equation}
With the convention $e^{iS}$, the ordered vertex contribution is
$i\mathcal K_{F^3}(1,2,3)$.  The full cubic vertex is obtained by summing the
ordered expression over the external-leg permutations, in direct analogy with
the Yang--Mills vertex in the main text.  Since the argument below applies to
each ordering separately, it fixes the vanishing of the complete amplitude.

Suppose first that the local coefficient is nonzero.  Then
$\sigma_1=\sigma_2=\sigma_3\equiv\sigma$, and therefore
\begin{equation}
 \epsilon_r=\sigma\lambda_r,
 \qquad
 K=\sigma\sum_{r=1}^3\lambda_r k_r
 =2\sigma\lambda_k,
 \qquad
 \lambda_k=\frac12\sum_{r=1}^3\lambda_r k_r.
 \label{eq:appB-K-lambdak}
\end{equation}
The only potentially nonzero time channels
Eq.~\eqref{eq:appB-time-channels} consequently require
\begin{equation}
 \lambda_k=0,\pm1.
 \label{eq:appB-special-lambdak}
\end{equation}
The spatial intertwiner, on the other hand, has the representation-theoretic
form
\begin{align}
 I_3\mathrel{\propto}{}&
 \mathsf S\sqrt{\lambda_k^2-1}
 \begin{pmatrix}j_1&j_2&j_3\\
 \epsilon_1\nu_1&\epsilon_2\nu_2&\epsilon_3\nu_3\end{pmatrix}
 \nonumber\\
 &\times
 \begin{pmatrix}j'_1&j'_2&j'_3\\
 \epsilon_1\nu'_1&\epsilon_2\nu'_2&\epsilon_3\nu'_3\end{pmatrix},
 \label{eq:appB-general-I3}
\end{align}
with the phases and square-root branch specified in
Eqs.~\eqref{eq:appA-outgoing-total-phase} and
\eqref{eq:appA-helicity-branch}.  The values $\lambda_k=\pm1$ vanish through
the explicit square root.  The remaining possibility $\lambda_k=0$ is
incompatible with the two triangle conditions.  Indeed, equal helicities
cannot give $\lambda_k=0$ because every $k_r$ is positive.  If two helicities
are positive and the third is negative, relabeling the negative-helicity mode
as leg 3 gives $k_3=k_1+k_2$ and
\begin{equation}
 j_3=\frac{k_1+k_2}{2}
 >j_1+j_2=\frac{k_1+k_2}{2}-2.
 \label{eq:appB-zero-triangle}
\end{equation}
This violates the unprimed triangle condition.  With two negative helicities
and one positive helicity, we instead relabel the positive-helicity mode as
leg 3.  The same argument applied to the primed spins then gives
$j'_3>j'_1+j'_2$.  Thus $I_3=0$ in every potentially nonzero time channel, and
\begin{equation}
 \mathcal C(\epsilon_r,\lambda_r)\,
 {\cal I}_K\,I_3=0
 \label{eq:appB-no-overlap}
\end{equation}
for every physical triple.  If the chirality condition fails, the local
coefficient vanishes.  If it holds, a nonzero time factor multiplies a zero
spatial intertwiner, while a nonzero spatial intertwiner multiplies a zero time
factor.  This is the promised no-overlap mechanism.

\subsection{Explicit checks}

As a check of the analytic argument, we evaluated
Eqs.~\eqref{eq:appB-local-contraction}, \eqref{eq:appB-time-result}, and
\eqref{eq:appB-no-overlap} for 150 exact mode configurations.  The sample
covers all eight helicity patterns and all eight incoming/outgoing patterns.
The configurations fall into three complementary classes:
\begin{equation}
 \begin{array}{c|c|c|c}
 \text{number}&\mathcal C&I_3&{\cal I}_K\\ \hline
 89&0&\neq0&\text{arbitrary}\\
 31&\neq0&\neq0&0\\
 30&\neq0&0&\neq0\quad(K=0,\pm2)
 \end{array}.
 \label{eq:appB-check-table}
\end{equation}
The first class tests the local chirality rule, the second tests the vanishing
time factor when the spatial overlap is nonzero, and the third tests the
vanishing spatial overlap in the special channels $K=0,\pm2$.  In every case,
the Lorentz-index contraction agrees exactly with $k_1k_2k_3\mathcal C$ times
the triple wedge product, while direct time integration agrees with
Eq.~\eqref{eq:appB-time-result}.  All 150 ordered kernels vanish.


\begin{thebibliography}{99}
\bibitem{Taylor:2017sph}
T. R. Taylor,
``A Course in Amplitudes,''
Phys. Rept. \textbf{691}, 1-37 (2017)
doi:10.1016/j.physrep.2017.05.002
[arXiv:1703.05670 [hep-th]].
\bibitem{Taylor:2024vdc}
T.~R.~Taylor and B.~Zhu,
``Scattering of quantum particles in global de Sitter spacetime I: The formalism,''
Nucl. Phys. B \textbf{1018}, 116999 (2025)
doi:10.1016/j.nuclphysb.2025.116999
[arXiv:2411.02504 [hep-th]].
\bibitem{Taylor:2025deepIR}
T.~R.~Taylor and B.~Zhu,
``Scattering of quantum particles in global de Sitter spacetime II: Scalars in deep infrared,''
Nucl. Phys. B \textbf{1022}, 117265 (2026)
doi:10.1016/j.nuclphysb.2025.117265
[arXiv:2509.25407 [hep-th]].
\bibitem{Marolf:2012kh}
D.~Marolf, I.~A.~Morrison and M.~Srednicki,
``Perturbative S-matrix for massive scalar fields in global de Sitter space,''
Class. Quant. Grav. \textbf{30}, 155023 (2013)
doi:10.1088/0264-9381/30/15/155023
[arXiv:1209.6039 [hep-th]].
\bibitem{Melville:2024ove}
S.~Melville and G.~L.~Pimentel,
``A de Sitter S-matrix from amputated cosmological correlators,''
JHEP \textbf{08}, 211 (2024)
doi:10.1007/JHEP08(2024)211
[arXiv:2404.05712 [hep-th]].
\bibitem{Dixmier:1961fbm}
J.~Dixmier,
``Repr{\'e}sentations int{\'e}grables du groupe de De Sitter,''
Bull. Soc. Math. Fr. \textbf{79}, 9-41 (1961)
doi:10.24033/bsmf.1558\ .
\bibitem{rep4}
J.~Penedones, K.~Salehi Vaziri and Z.~Sun,
``Hilbert space of quantum field theory in de Sitter spacetime,''
Phys. Rev. D \textbf{111}, no.4, 045001 (2025)
doi:10.1103/PhysRevD.111.045001
[arXiv:2301.04146 [hep-th]].
\bibitem{Lindsay:2025chz}
A.~Lindsay and T.~R.~Taylor,
``Symmetries of de Sitter particles and amplitudes,''
JHEP \textbf{04}, 064 (2026)
doi:10.1007/JHEP04(2026)064
[arXiv:2512.13781 [hep-th]].
\bibitem{rep21}
A.~Rios Fukelman, M.~Semp{\'e} and G.~A.~Silva,
``Notes on gauge fields and discrete series representations in de Sitter spacetimes,''
JHEP \textbf{01}, 011 (2024)
doi:10.1007/JHEP01(2024)011
[arXiv:2310.14955 [hep-th]].
\bibitem{Lehoucq:2002E}
R.~Lehoucq, J.~P.~Uzan and J.~Weeks,
``Eigenmodes of lens and prism spaces,''
Kodai Math. J. \textbf{26}, 119-136 (2003)
doi:10.48550/arXiv.math/0202072
[arXiv:math/0202072 [math.SP]].
\bibitem{BenAchour:2015aah}
J.~Ben Achour, E.~Huguet, J.~Queva and J.~Renaud,
``Explicit vector spherical harmonics on the 3-sphere,''
J. Math. Phys. \textbf{57}, no.2, 023504 (2016)
doi:10.1063/1.4940134
[arXiv:1505.03426 [math-ph]].
\bibitem{varsh} D. A. Varshalovich, A. N. Moskalev, and V. K. Khersonskii, ``Quantum Theory of Angular Momentum,'' World Scientific Publishing Co. Inc., Singapore (1988).
\bibitem{graph} E. Balcar and S.W. Lovesey, ``Introduction to the Graphical Theory of Angular Momentum,'' Springer-Verlag (2009).
\bibitem{Maldacena:2002vr}
J.~M.~Maldacena,
``Non-Gaussian features of primordial fluctuations in single field inflationary models,''
JHEP \textbf{05}, 013 (2003)
doi:10.1088/1126-6708/2003/05/013
[arXiv:astro-ph/0210603 [astro-ph]].
\bibitem{Maldacena:2011nz}
J.~M.~Maldacena and G.~L.~Pimentel,
``On graviton non-Gaussianities during inflation,''
JHEP \textbf{09}, 045 (2011)
doi:10.1007/JHEP09(2011)045
[arXiv:1104.2846 [hep-th]].
\bibitem{Arkani-Hamed:2015bza}
N.~Arkani-Hamed and J.~Maldacena,
``Cosmological Collider Physics,''
[arXiv:1503.08043 [hep-th]]
\bibitem{Chen:2017ryl}
X.~Chen, Y.~Wang and Z.~Z.~Xianyu,
``Schwinger-Keldysh Diagrammatics for Primordial Perturbations,''
JCAP \textbf{12}, 006 (2017)
doi:10.1088/1475-7516/2017/12/006
[arXiv:1703.10166 [hep-th]].
\bibitem{MdAbhishek:2025dhx}
M.~Abhishek, C.~Sleight and M.~Taronna,
``Cosmological Correlators in Gauge Theory and Gravity from EAdS,''
[arXiv:2509.09536 [hep-th]].
\bibitem{Arundine:2026fbr}
M.~Arundine, D.~Baumann, M.~H.~G.~Lee, G.~L.~Pimentel and F.~Rost,
``The Cosmological Grassmannian,''
[arXiv:2602.07117 [hep-th]].
\bibitem{six} Akshay Venkatesh and X. Griffin Wang, ``The Tetrahedral (or $6\mathsf{j}$) Symbol,'' [arXiv:2602.14908 [math.NT]].\
\bibitem{Strominger:2021mtt}
A.~Strominger,
``$w_{1+\infty}$ Algebra and the Celestial Sphere: Infinite Towers of Soft Graviton, Photon, and Gluon Symmetries,''
Phys. Rev. Lett. \textbf{127}, no.22, 221601 (2021)
doi:10.1103/PhysRevLett.127.221601
[arXiv:2105.14346 [hep-th]].
\bibitem{Taylor:2023ajd}
T.~R.~Taylor and B.~Zhu,
``$w_{1+\infty}$ Algebra with a Cosmological Constant and the Celestial Sphere,''
Phys. Rev. Lett. \textbf{132}, no.22, 221602 (2024)
doi:10.1103/PhysRevLett.132.221602
[arXiv:2312.00876 [hep-th]].
\bibitem{Bittleston:2024rqe}
R.~Bittleston, G.~Bogna, S.~Heuveline, A.~Kmec, L.~Mason and D.~Skinner,
``On AdS$_{4}$ deformations of celestial symmetries,''
JHEP \textbf{07}, 010 (2024)
doi:10.1007/JHEP07(2024)010
[arXiv:2403.18011 [hep-th]].
\bibitem{Basile:2024ydc}
T.~Basile, E.~Joung, K.~Mkrtchyan and M.~Mojaza,
``Spinor-helicity representations of particles of any mass in dS4 and AdS4 spacetimes,''
Phys. Rev. D \textbf{109}, no.12, 125003 (2024)
doi:10.1103/PhysRevD.109.125003
[arXiv:2401.02007 [hep-th]].
\end{thebibliography}
\end{document}